\documentclass[pre,twocolumn,superscriptaddress,nofootinbib,amsmath,amssymb,longbibliography]{revtex4-1}
\usepackage{graphicx}
\usepackage{float}
\usepackage[usenames,dvipsnames]{xcolor}
\usepackage[colorlinks=true,citecolor=Blue,linkcolor=RubineRed,urlcolor=Blue]{hyperref}
\begin{document}
\title{Reversible, Irreversible and Mixed Regimes for
Periodically Driven Disks in Random Obstacle Arrays
} 
\author{D. Minogue}
\affiliation{Department of Physics, University of Notre Dame, Notre Dame, Indiana 46656 USA}
\author{M. R. Eskildsen}
\affiliation{Department of Physics, University of Notre Dame, Notre Dame, Indiana 46656 USA}
\author{C. Reichhardt}
\affiliation{Theoretical Division and Center for Nonlinear Studies,
Los Alamos National Laboratory, Los Alamos, New Mexico 87545, USA}
\author{C. J. O. Reichhardt}
\affiliation{Theoretical Division and Center for Nonlinear Studies,
Los Alamos National Laboratory, Los Alamos, New Mexico 87545, USA}

\date{\today}

\begin{abstract}
 We examine an assembly of repulsive disks interacting with a random obstacle array under a periodic drive, and find a transition from reversible to irreversible dynamics as a function of drive amplitude or disk density. At low densities and drives, the system rapidly forms a reversible state where the disks return to their exact positions at the end of each cycle. In contrast, at high amplitudes or high densities, the system enters an irreversible state where the disks exhibit normal diffusion. Between these two regimes, there can be a glassy irreversible state where most of the system is reversible, but localized irreversible regions are present that are prevented from spreading through the system due to a screening effect from the obstacles. We also find states that we term combinatorial reversible states in which the disks return to their original positions after multiple driving cycles. In these states, individual disks exchange positions but form the same configurations during the subcycles of the larger reversible cycle. 
\end{abstract}
\maketitle

\section{Introduction}

There are a variety of systems that can be modeled as a collection
of particles driven over a disordered landscape, such as
vortices in type-II superconductors \cite{Bhattacharya93,Reichhardt17},
colloidal particles \cite{Pertsinidis08}, magnetic skyrmions
\cite{Jiang17,Reichhardt22a}, emulsions \cite{LeBlay20},
active matter \cite{Morin17,Sandor17a}, and granular matter \cite{Reichhardt12}.
The quenched disorder can be in the form of pinning sites that act as
local traps, such as those found in superconducting vortex systems,
or of obstacles, such as those found in soft matter systems.
Other examples of this type of dynamics include particulate matter
flowing through disordered media or bottlenecks,
where clogging phenomena can occur
\cite{Zuriguel15,Nguyen17,Gerber18,Stoop18}.
In systems with pinning or obstacles, the drive
responsible for producing flow is generally
applied only along one direction;
however, in some situations the drive is oscillating,
and in this case, it is possible for reversible motion to appear in which
the particles return to the same positions
at the end of each drive cycle or after a
fixed number of cycles \cite{Reichhardt23}.

Reversible to irreversible (R-IR) transitions in the absence of pinning
or obstacles have been studied in a variety of systems.
One of the simplest of these systems is 
dilute suspensions of colloids under a periodic shear,
where 
it was shown that for a fixed colloid density,
there is a critical shear amplitude below which the system
organizes to a reversible state, while at high amplitude, the system
remains in a chaotic state where the particles exhibit diffusive behavior
\cite{Pine05,Corte08}.
If the drive amplitude is fixed, there is also a critical density
below which the
system forms a reversible state.
Similar R-IR transitions have been studied in
other periodically sheared dilute systems where,
in some cases, the reversible states were found to exhibit hyperuniformity
\cite{Hexner15,Tjhung15,Weijs15,Reichhardt19,Lei19a}.
For the dilute system, the reversible states are usually
those in which collisions between the particles no longer occur,
and such states have been shown to be capable of
encoding memories of the number of cycles through which the system passed
on the way to the reversible state
\cite{Paulsen14,Keim19}.
R-IR transitions
have also been studied in dense systems where 
the particles are in continuous contact, such as granular
matter \cite{Schreck13,Royer15} or amorphous solids.
In this case, the reversible state of the system
is marked by reversible plastic events
\cite{Regev13,Fiocco13,Fiocco14,Keim14,Priezjev16,Priezjev17}.
Dense systems can also show reversibility after
multiple cycles due to the appearance of multiple
plastic events that interact by long-range strain
fields \cite{Regev13,Lavrentovich17,Khirallah21,Keim21}.

R-IR transitions can also occur 
in systems that exhibit pinning or clogging dynamics,
where the cyclically driven  particles interact  
with quenched disorder  \cite{Reichhardt17}.
Such systems include
vortices in type-II superconductors
\cite{Mangan08,Okuma11,Pasquini21,Maegochi19,Maegochi21},
magnetic skyrmions \cite{Litzius17,Brown18}, and
colloidal particles \cite{Stoop18}. 
One difference between systems with and without quenched disorder is
that when quenched disorder is present, R-IR transitions can be induced
with a uniformly applied drive rather than through shearing;
however, under uniform driving
in the absence of quenched disorder or thermal fluctuations, only
reversible states form.
Systems with quenched disorder can behave elastically,
where particles maintain the same nearest neighbors over time, or
plastically, where deformations cause particles to exchange neighbors
or lead to the coexistence of
flowing and pinned states \cite{Reichhardt17}. 

In cyclically-driven superconducting vortex systems,
plastic deformations were shown to result in the
particles undergoing chaotic motion,
while when the drive or pinning is weak, the system can form reversible orbits
\cite{Mangan08,Okuma11,Pasquini21,Maegochi19,Maegochi21}. 
In superconducting vortex and magnetic skyrmion systems,
the quenched disorder takes the form of randomly located trapping
sites; however, there have also been studies of R-IR transitions
in cyclically-driven disk systems interacting
with a periodic array of obstacles \cite{Reichhardt22}.
For the latter case, when the driving is applied along a
symmetry direction of the obstacle array,
the system forms reversible and spatially ordered states;
however, for drives applied along angles that are incommensurate
with the array symmetry, the system forms an irreversible state
even at low drives. Stoop {\it et al.} \cite{Stoop18} also considered
disks moving over a random array of obstacles
under a forward and backward pulse drive,
and found that the system can form a partially clogged state
with different configurations during different portions of the drive.
This work suggests that R-IR transitions could
also be possible in disordered obstacle arrays.

Here, we consider a two-dimensional assembly of disks cyclically driven
over a random array of obstacles.
We find that for high drive amplitudes or high disk densities,
the system forms irreversible states with diffusive behavior, while
for lower drives and densities, reversible states
occur that return to the original configuration
after one or more drive cycle. We map out the onset of the R-IR
transition as a function of disk density and drive amplitude.
In some cases, the reversible states
consist of clogged regions that coexist with regions of moving disks,
while the irreversible states can also form heterogeneous configurations
that change from one cycle to the next.
Near the R-IR boundary, we find what we call glassy irreversible states
where most of the system
is reversible, but there are localized irreversible regions that are
screened from rapidly spreading
through the system by a trapping effect of the obstacles.
This leads to extended times during which the system behaves
subdiffusively.
The localized chaotic regions slowly move through the system after
many cycles.
We also find states that do not have long-time diffusion but
contain small chaotic regions that remain localized.
We observe a number of states that are reversible after multiple cycles,
and term these combinatorial reversible states. They are
associated with groups of disks that exchange
positions such that after $N$ cycles, the macroscopic disk configuration
is the same but the microscopic positions of the disks differs.
The system returns to the exact same configuration of the
original disks after multiples of the $N$ cycles.
These combinatorial multi-cycle states are
distinct from the multiple-cycle states found in amorphous solids,
which occur due to longer-range elastic interactions. In our
disk system, they occur due to purely local contact interactions.

\section{Simulation}

We examine a two-dimensional system
of size $L \times L$
containing $N_d$ mobile disks of radius $r_d=0.5$
and $N_{\rm obs}$ obstacles of radius $r_{\rm obs}=1.025$.
The sample has periodic boundary conditions in the $x$ and $y$ directions.
The density $\rho$ is
defined to be the area covered by the obstacles and mobile disks,
$\rho = N_{\rm obs}\pi r^2_{\rm obs}/L^2 + N_{d}\pi r^2_{d}/L^2$, where
we fix $L = 36$.
We also fix the number of obstacles to $N_{\rm obs}=80$.
The dynamics of the mobile disks is obtained from
the following overdamped equation of motion:
\begin{equation} 
\alpha_d {\bf v}_{i}  =
{\bf F}^{\rm dd}_{i} +  {\bf F}^{\rm obs}_{i} + {\bf F}^{\rm D}  \ .
\end{equation}
Here, $\alpha_d$ is the damping constant, which we set to unity. The disk
velocity is ${\bf v}_{i} = d {\bf r}_{i}/dt$ where ${\bf r}_i$ is
the location of disk $i$.
The disk-disk interaction force
${\bf F}^{\rm dd}_{i} = \sum_{j \neq i}^{N_d} k (D - r_{ij}) \Theta(D - r_{ij}){\bf \hat{r}}_{ij}$
is represented by a short-range harmonic repulsive potential,
where $D=2r_d$, $k$ is the spring constant,
$r_{ij}=|{\bf r}_i-{\bf r}_j|$, and
${\bf \hat{r}}_{ij}=({\bf r}_i-{\bf r}_j)/r_{ij}$.
The disk-obstacle interaction term ${\bf F}^{\rm obs}_i$ has the same
form as ${\bf F}^{\rm dd}$ but with $D=r_d+r_{\rm obs}$,
so that the obstacles are represented as
randomly located non-overlapping immobile disks of radius $r_{\rm obs}$.
The driving force ${\bf F}^{\rm D}=\pm A {\bf \hat{x}}$
is a zero-centered square wave of amplitude $A$ and
period $T=4 \times 10^5$ simulation time steps,
where the positive sign is used during the first half of
each period and the negative sign is used during the second half period.
As a measurement of time we use the quantity $N_c$, which is the total
number of driving cycles that have elapsed since the beginning of the
simulation.
We hold $T$ fixed throughout this work and vary $A$ over the range
$A=0.01$ to $A=0.05$.
To initialize the system we place the disks in randomly chosen locations such
that each disk does not overlap with any other disks or with any obstacles.

To quantify the number of disks that
return to their original positions after $n$ driving cycles, we
measure the difference between 
the disk positions at a reference time $t_0$ and
a time $t_0 +nT$ that is exactly $n$ driving cycles later:
$R(n) = \sum^{N_{d}}_{i}[{\bf r}_{i}(t_{0} + nT) - {\bf r}_{i}(t_{0})]$.
When $R(n)=0$, the motion is reversible after $n$ driving cycles.
If $R(1)=0$, then the motion is reversible after only a single
driving cycle.
The total distance traveled by the disks after $N_c$ cycles
have elapsed from a reference time $t_0$
is given by
$d(N_c) = \sum^{N_d}_{i}|{\bf r}_i(t_0+N_c)-{\bf r}_i(t_0)|$.
In an irreversible state, $d(N_c)$ will grow continuously as a function
of $N_c$.
Since $N_c$ is proportional to time, we also characterize
irreversiblity by fitting $d \propto N_c^\alpha$,
where $\alpha=0$ for reversible motion,
$\alpha=1.0$ for Brownian diffusion,
and $0 < \alpha < 1$ for subdiffusion.
In each case, the results are averaged
over five different obstacle realizations.

\section{Results}

\begin{figure}
  \includegraphics{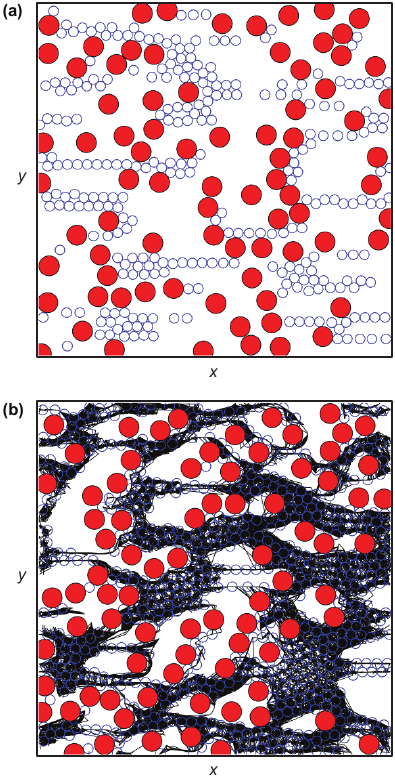}
\caption{Obstacle locations (red circles) and mobile disk locations (blue)
along with trajectory
lines indicating the net translation of the mobile disks from
one cycle to the next over a time of $N_c=100$ driving cycles.  
The disk positions are shown at the end of the drive cycle when the drive is
about to switch from the $-x$ direction back to the $+x$ direction.
(a) A completely reversible state at $A = 0.02$ and $\rho = 0.368$,
where all of the disks return to the same position at the end of
each driving cycle. The trajectory lines have zero length and thus do
not appear in the panel.
(b) An irreversible state at $A  = 0.015$ and $\rho = 0.579$.
The trajectory lines are finite and disordered.
}
\label{fig:1}
\end{figure}

In Fig.~\ref{fig:1}(a) we show a snapshot of
the obstacle and disk positions in the
completely reversible state at
$A = 0.02$ and $\rho = 0.368$.
The snapshot is taken at the end of the drive cycle when the drive is
about to switch from the $-x$ direction back to the $+x$ direction.
We find that a reversible state can appear
even when some disks come into contact
with other disks, in contrast to the sheared dilute disk systems,
where the reversible states involve no disk collisions.
Some of the disks become clogged in bottleneck configurations during
portions of the drive cycle, and these stuck regions coexist with other
regions where the disks continue to move throughout the cycle. When the
drive direction switches, the bottleneck regions are released and become
mobile again, but new bottlenecks can form in different locations for the
reversed driving direction.
In Fig.~1 we plot the trajectories of the disks showing the motion from the
end of one drive cycle to the end of the next drive cycle during $N_c=100$
cycles,
but in the reversible state of Fig.~\ref{fig:1}(a), these trajectories are
of zero length and do not appear in the panel.
Figure~\ref{fig:1}(b) shows an irreversible state at $A  = 0.015$
and $\rho = 0.579$.
Here, almost all of the disks participate in the irreversible behavior.
The system still forms a heterogeneous state in which temporarily clogged
and flowing regions coexist as a function of time, but the arrangement and
location of the clogged regions changes over time instead of reaching a
permanent repeating cycle.

\begin{figure}
  \includegraphics[width=\columnwidth]{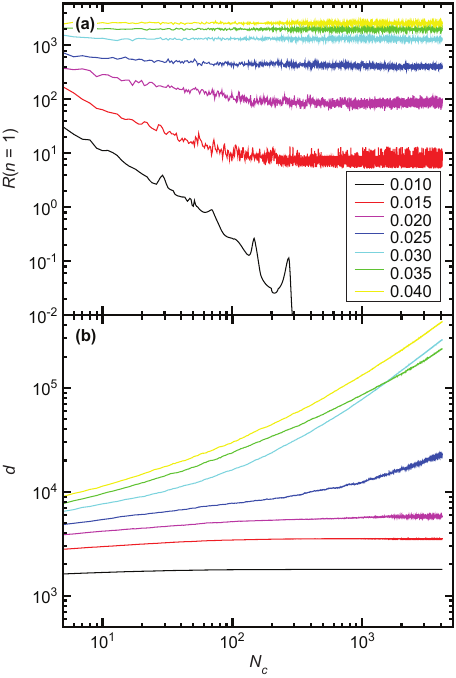}
\caption{
A sample with total disk density $\rho=0.398$ at varied drive
amplitudes $A=0.01$ to $A=0.04$.
(a) The distance $R(n=1)$ traveled by the disks in a single driving cycle
vs the total number of elapsed driving cycles $N_c$, where we have
set $t_0=N_c-1$ in the calculation of $R$.
(b) The total distance traveled $d$ vs $N_c$.
For $A = 0.01$, 0.015, and $0.02$, the system reaches
a reversible state. For $A = 0.025$, the system is irreversible
but shows subdiffusive behavior, and for
$A = 0.03$, 0.035, and $0.04$, the motion is 
irreversible with regular diffusion.
}
\label{fig:2}
\end{figure}

In Fig.~\ref{fig:2}(a),
for a sample with $\rho=0.398$ under different drive amplitudes $A$,
we plot $R(n=1)$, the distance the disks
travel during a single cycle, as a function of the total number $N_c$
of elapsed cycles. Here we set $t_0=N_c-1$ in the calculation of
$R$.
Figure~\ref{fig:2}(b) shows the corresponding total distance
traveled $d$ versus $N_c$.
For $A = 0.01$, 0.015, and $0.02$,
the system reaches a reversible state in which
$d$ saturates to a constant value.
When $A  = 0.01$, in the steady state the system
is reversible after one cycle,
so $R(n=1)$ drops to zero in Fig.~\ref{fig:2}(a),
while for $A = 0.015$ and $0.02$, the steady state positions
recur only after multiple cycles, so the $R(n=1)$ curve
saturates to a finite value.
For $A= 0.025$
the system is irreversible, but during an extended period of time
$d$ grows less than linearly with time, indicating
subdiffusive behavior.
In the irreversible states
at $A = 0.03$, $0.035$, and $0.04$, $d$ grows linearly
with time, a signature of Brownian diffusion.

\begin{figure}
  \includegraphics[width=\columnwidth]{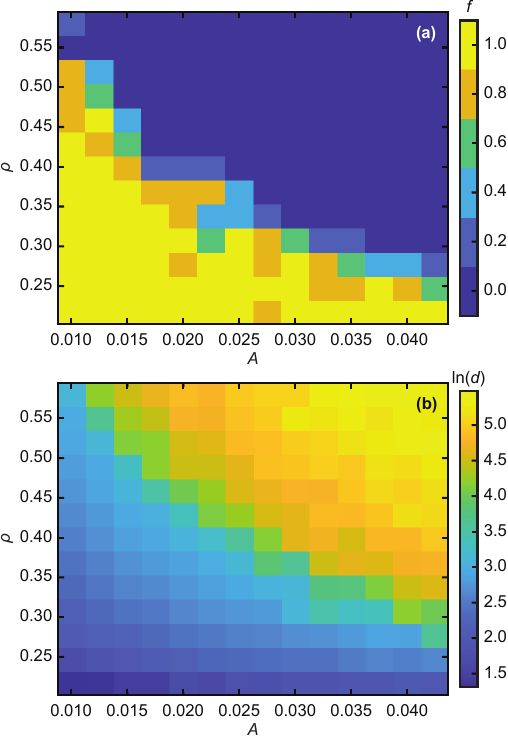}
\caption{
(a) Heat map of the fraction $f$ of the five different disorder realizations
that reached reversible states plotted as a function of $\rho$ vs $A$. 
(b) Heat map of $\ln[d(N_c=3000)]$, the total displacement measured
during $N_c=3000$ driving cycles with $t_0=1000$ driving cycles,
as a function of $\rho$ vs $A$.
}
\label{fig:3}
\end{figure}

In Fig.~\ref{fig:3}(a), we plot a heat map showing the fraction $f$ of the
five different disorder realizations that reached reversible states as
a function of density $\rho$ versus drive amplitude $A$.
As $\rho$ decreases, the threshold value of $A$ at which irreversible
behavior disappears shifts upward, so that reversible states appear for
small $\rho$ and $A$ while irreversible states appear for large $\rho$ and
$A$.
The boundary separating reversible and irreversible states is not
sharp; instead, the
disorder realizations are split with a portion of the realizations
becoming reversible and the remaining realizations remaining irreversible.
In our study, we focused on densities less than $\rho=0.6$; however,
for higher densities,
the system could become completely clogged or mostly clogged, in which case
the fraction of reversible states could increase again.
In Fig.~\ref{fig:3}(b) we plot
a heat map of the logarithm of $d(N_c=3000)$, measured 
using $t_0=1000$ driving cycles, as a function of $\rho$ versus $A$. 
Near the crossover from reversible to irreversible behavior,
the total displacement begins to increase significantly.

\begin{figure}
  \includegraphics[width=\columnwidth]{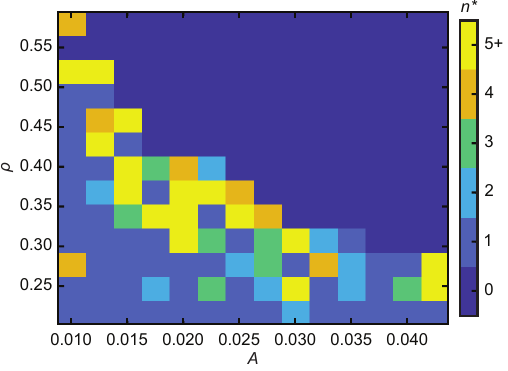}
\caption{
Heat map of $n^*$, the smallest value of $n$ for which $R(n)$ reaches zero
in the reversible regime, as a function of $\rho$ vs $A$.
When $n^*>1$, the state is multi-cycle reversible.
In the irreversible
regime, we mark $n^*=0$. In the reversible regime, the value of $n^*$ is
averaged only over disorder realizations that were reversible.
}
\label{fig:4}
\end{figure}

By measuring $R(n)$ for different values of $n$ in the reversible regime,
we can determine how many drive cycles are required for the disks to
reach their original positions. The smallest value of $n$ for which
$R(n)$ reaches zero in the reversible regime is labeled $n^{*}$.
In Fig.~\ref{fig:4}, we plot a heat map of $n^*$ as a function
of $\rho$ versus $A$. In the reversible regime, we average $n^*$ only over
the disorder realizations that were reversible, while in the irreversible
regime where the measure is not defined, we mark $n^*=0$.
For low $\rho$ and low $A$, we primarily find $n^*=1$, meaning that most
of the states are reversible after one cycle, while near the crossover from
reversible to irreversible behavior, we find multi-cycle reversible states
with $n^*=5$ or more, and even observed one state that was reversible
after $n^*=24$ drive cycles.

\begin{figure}
  \includegraphics[width=\columnwidth]{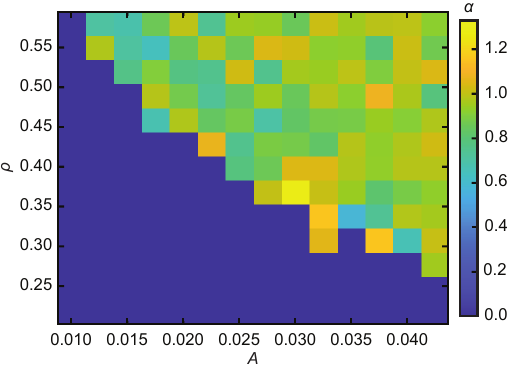}
\caption{
Heat map as a function of $\rho$ vs $A$
of the exponent $\alpha$ obtained from a fit to $d \propto N_c^\alpha$.
$\alpha=0$ indicates reversible behavior, and $\alpha=1.0$ indicates diffusive
behavior.
}
\label{fig:5}
\end{figure}

\begin{figure}
  \includegraphics{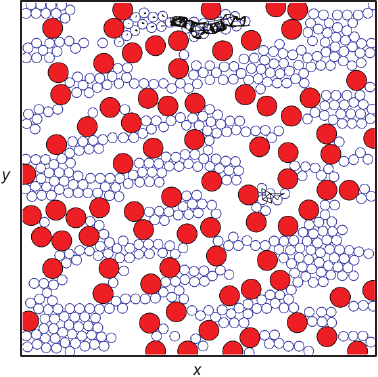}
\caption{
Obstacle locations (red circles) and mobile disk locations (blue) along with
trajectory lines indicating the net translation of the mobile disks from
one cycle to the next over a time of $N_c=100$ driving cycles in
a state with two local regions of irreversible behavior
but no long time diffusion at $\rho=0.519$ and $A=0.01$.
}
\label{fig:6}
\end{figure}

In Fig.~\ref{fig:5}, we plot a heat map
as a function of
$\rho$ versus $A$ of the exponent $\alpha$
obtained from fits to
$d \propto N_c^\alpha$.
In a reversible state, $\alpha=0$, while a value of $\alpha=1.0$ indicates
Brownian motion.
We find regions in which $0 < \alpha < 1$, indicating that subdiffusion
is occurring over an extended number of drive cycles.
In these instances, the system forms what we
call a glassy irreversible state where large
sections of the system act reversibly,
but there are localized regions in which chaotic motion occurs.
These localized regions can slowly move through the sample
after many cycles.
Such glassy reversible states,
where there are localized regions of
irreversible behavior that coexist with reversible regions,
are likely produced by a screening effect from the obstacles,
which prevents irreversible regions from making contact with spatially
separated reversible regions of the sample.
In the irreversible regime of sheared systems without obstacles,
no such screening exists and the irreversible motion can spread
unhindered throughout the sample.   

We have also found states that show local irreversibility
but do not exhibit long-time diffusion, again due to a screening effect
of the obstacles. For example, if obstacles completely surround a
region of disks, this region can undergo irreversible or chaotic motion
that is effectively trapped and cannot interact with
other parts of the system.
These disks can continuously change their configurations,
so their behavior is irreversible,
but the confinement effect limits the maximum distance they can travel
and bounds the maximum possible diffusion.
In this way, a portion of the system would be locally ergodic,
but the overall system is not globally ergodic. If the confined
region is sufficiently small, the disks may be able to regain their original
positions eventually, but they will not repeatedly return to these original
positions in a periodic manner, so they will never enter a multi-cycle
reversible state.
Figure~\ref{fig:6} shows
an example of this behavior at $A = 0.01$ and $\rho = 0.519$,
where there is no long-time diffusion, but there are
two regions that are locally chaotic.

\section{Multi-Cycle Combinatorial Reversible States}

In previous work in dense amorphous systems,
multi-cycle reversible states were observed in which
the particles form complex loop-like orbits that return to
the same point after $N=n^*$ cycles
\cite{Regev13,Lavrentovich17,Khirallah21,Keim21}.
In these systems, the particle orbits are different during
each of the $N$ driving cycles and only repeat once the entire
cycle has been completed.
Additionally, multi-cycle states are linked to the occurrence of
distinct plastic events that can interact with each other
through a long-range strain field.
In our system, the reversible multi-cycle state
is associated with groups of particles which can adopt the same macroscopic
configuration multiple times during the full cycle, but which only reach the
original microscopic disk configuration after the cycle is complete,
creating what we call a combinatorial reversible state.

\begin{figure}
  \includegraphics[width=\columnwidth]{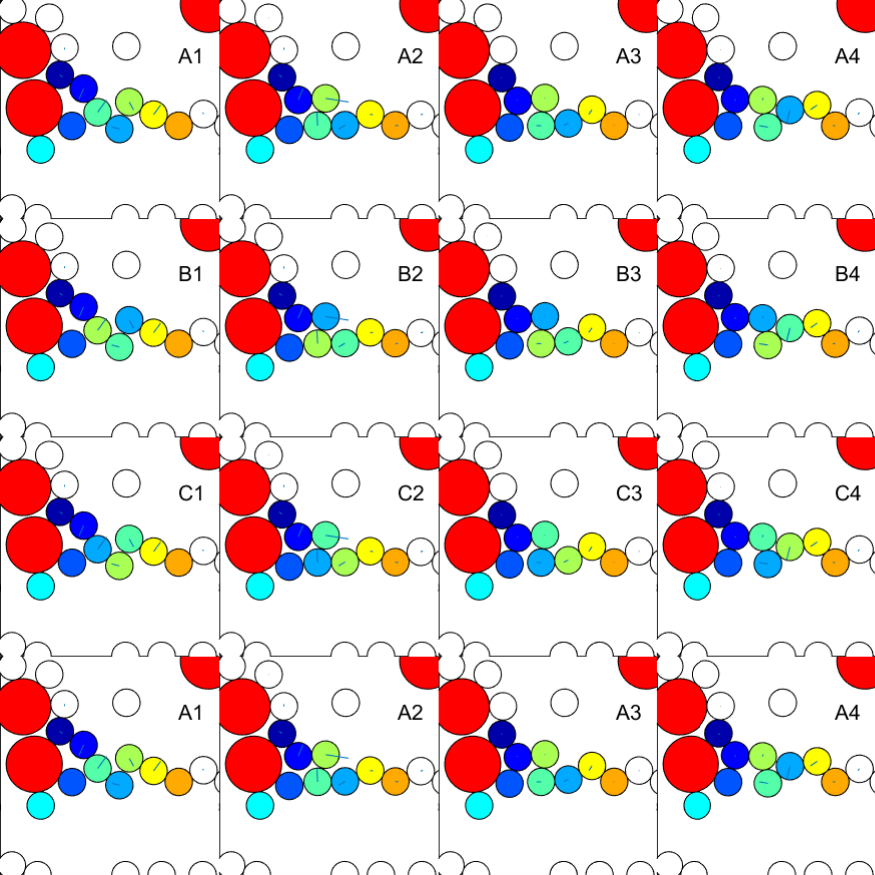}
\caption{
Illustration of a combinatorial reversible state at $\rho=0.337$ and $A=0.025$
where the obstacle locations (large red circles),
nonexchanging mobile disks (white
circles), and exchanging mobile disks (small colored circles) are shown in
only a small portion of the sample.
The disks return to their original positions every twelve cycles.
Each panel shows the disk configuration at the end of a drive cycle.
Time increases from left to right and top to bottom so that the
cycle sequence is A1-A2-A3-A4-B1-B2-B3-B4-C1-C2-C3-C4. The
macroscopic disk
configurations repeat every four drive cycles, so that A1, B1, and
C1 have the same macroscopic disk configuration,
but the individual disks are
permuted within this configuration.
The small trajectory lines indicate the net distance moved by each disk
compared to its position at the end of the previous driving cycle.
}
\label{fig:7}
\end{figure}

In Fig.~\ref{fig:7}, we show an example of a combinatorial
reversible state where we highlight the positions of
14 disks and three obstacles in a small portion of a sample with
$\rho=0.337$ and $A=0.025$ that is multi-cycle reversible.
The white disks return to their original positions after every cycle,
and we give distinct colors to
the nine disks that reach different positions from cycle
to cycle but only return to their original positions after twelve
cycles.
The trajectory lines connect the starting point of the disk from the end of
the previous drive cycle to its ending point at the end of the illustrated
drive cycle.
It is important to remember that in between the snapshots shown in
each panel of Fig.~\ref{fig:7}, the disks move back and forth through a
complete driving cycle, so that although their net motion is small, their
actual motion is not small.
Panel A1 shows the starting configuration.
The sample progresses
through configurations A2, A3, and A4, and after a total of four drive
cycles, the macroscopic disk configuration in panel B1 is exactly the same
as that of panel A1. The individual disk positions are not the same, however;
the blue, green, and blue-green disks at the center of the image have
exchanged places. After four more cycles, the system has passed through
states B2, B3, and B4, and reached configuration C1. This is again
macroscopically the same as A1 but microscopically different, with the
blue, green, and blue-green disks having rotated into yet another
arrangement. Four cycles later, the system passes through C2, C3, and C4,
and reaches the original state A1. Similar combinatoric swaps separate the
macroscopically identical but microscopically distinct states A2, B2, and
C2. The same is true for states A3, B3, and C3 as well as states
A4, B4, and C4.
In this way, the disks are {\it fully} reversible
after 12 cycles, but their macroscopic
{\it configuration} is reversible every four cycles.
This particular region of the sample acts like a small rotating gear.

\begin{figure}
  \includegraphics[width=\columnwidth]{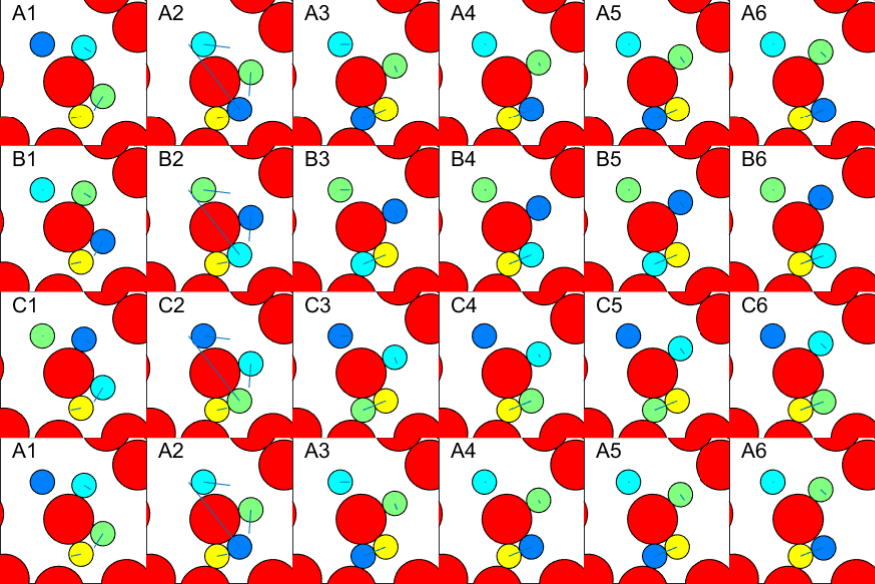}
\caption{
Illustration of a different small 
region of the combinatorial reversible state
from Fig.~\ref{fig:7} at $\rho=0.337$ and $A=0.025$, showing the
obstacle locations (large red circles) and exchanging mobile disks (small
colored circles).
The disks return to their original positions every 18 cycles.
Each panel shows the disk configuration at the end of a drive cycle.
Time increases from left to right and top to bottom.
The macroscopic disk configurations repeat every six drive cycles,
so that A1, B1, and C1 have the same macroscopic disk configuration, but the
individual disks are permuted within this configuration. The small
trajectory lines indicate the net distance moved by each disk compared to
its position at the end of the previous driving cycle.
}
\label{fig:8}
\end{figure}

In Fig.~\ref{fig:8}, we illustrate a different small portion of the
sample from Fig.~\ref{fig:7}. Here the disks return to their
original positions every 18 drive cycles but the macroscopic disk
configuration repeats every six drive cycles.
The lines highlight the net motion of the particles from their locations at
the end of the previous driving cycle.
To understand the motion of the disks, it is even more important to keep
in mind the fact that the disks translate through an entire drive cycle
in between consecutive frames of the figure. As a result, rather than the
relatively simple rotation illustrated in Fig.~\ref{fig:7}, we find in
Fig.~\ref{fig:8} that the disks can do a leapfrog position exchange.
The initial configuration is labeled A1. After one drive cycle, in A2,
the dark blue disk has interposed itself between the green and
yellow disks. Small adjustments of the disk positions occur during
cycles A3, A4, A5, and A6, until on the sixth cycle, in B1, the
macroscopic disk configurations of A1 are reproduced but with a permutation
in the disk positions. The same pattern repeats, with the light blue disk
interposing itself between the yellow and dark blue disks in panel B2,
followed by small disk adjustments for four cycles and a return to the A1
macroscopic configuration in the twelfth cycle, C1. After 18 cycles the
original disk configuration is restored. The leapfrog exchange does not
occur while the disks are surrounding the pictured obstacle; instead, it is
as the disks move during the driving cycle and make contact with other
obstacles (out of frame) and disks that their positions are swapped.
The localized nature of the
multi-cycle reversible states illustrated in Figs.~\ref{fig:7}
and ~\ref{fig:8} makes it possible for a single system to have numerous
multi-cycle states present simultaneously, so that the entire system becomes
fully reversible only after all of the multi-cycle states have reached their
starting configurations at the same time. In the case of the combination
shown, a 12-cycle reversible state with an 18-cycle reversible state, full
reversibility happens only after 36 cycles.
The number of possible multi-cycle regions increases as the boundary
between the reversible and irreversible behavior is approached,
and the necessity for simultaneous synchronization of multiple reversible
regions is responsible for
the large values of $n^*$
found near the reversible-irreversible boundary in Fig.~\ref{fig:4}.

\section{Summary}

We have examined the crossover from reversible to irreversible behavior in a
system of disks
moving through a random obstacle array 
under cyclic drive.
We measure the net displacement of the disks after $n$ cycles
for different disk densities and drive amplitudes.
For high densities and high amplitudes, we find
an irreversible state in which the disks undergo diffusive motion.
In the reversible state, for low densities and low amplitudes the
system returns to its original configuration after every drive cycle,
but as the reversible-irreversible boundary is approached,
multi-cycle reversible states appear in which the disks return to their
original configurations after two or more driving cycles.
We also observe multi-cycle combinatorial reversible states in which
the macroscopic disk configurations repeat after a subset of cycles but
the individual disk positions have been permuted, so that the original
positions are restored only after a sufficient number of permutation
cycles occur. This can produce very large multi-cycle reversibility when
more than one multi-cycle combinatorial region is present in the sample and
the regions do not have the same reversible period.
We find that some irreversible states
have what we call glassy irreversible properties, where regions
of disks exhibit chaotic irreversible behavior that
remains localized for long times
due to a screening effect from the obstacles.
In the glassy state, these irreversible regions gradually move
around the system.
In other cases, the localized irreversible regions become completely trapped,
so there is no long time diffusion in the system even though the behavior
remains irreversible.
Our results show that disks driven through
obstacles have behaviors
similar to what is found for dilute 
sheared systems, where reversible orbits form when no collisions occur between
the particles, as well as behaviors similar to what is
observed in
sheared dense amorphous systems,
where interactions between reversible regions can produce
multi-cycle reversibility.

\begin{acknowledgments}
This research was supported in part by the Notre Dame Center for Research
Computing. Work at the University of Notre Dame (DM, MRE) was supported
by the US Department of Energy, Office of Basic Energy Sciences,
under Award No. DE-SC0005051.
We gratefully acknowledge the support of the US Department of
Energy through the LANL/LDRD program for this work.
The work at LANL was supported by the US Department of Energy through
the Los Alamos National Laboratory. Los Alamos National Laboratory is
operated by Triad National Security, LLC, for the National Nuclear
Security Administration of the US Department of Energy
(Contract No. 892333218NCA000001).
\end{acknowledgments}

\bibliography{mybib}

\begin{thebibliography}{42}%
\makeatletter
\providecommand \@ifxundefined [1]{%
 \@ifx{#1\undefined}
}%
\providecommand \@ifnum [1]{%
 \ifnum #1\expandafter \@firstoftwo
 \else \expandafter \@secondoftwo
 \fi
}%
\providecommand \@ifx [1]{%
 \ifx #1\expandafter \@firstoftwo
 \else \expandafter \@secondoftwo
 \fi
}%
\providecommand \natexlab [1]{#1}%
\providecommand \enquote  [1]{``#1''}%
\providecommand \bibnamefont  [1]{#1}%
\providecommand \bibfnamefont [1]{#1}%
\providecommand \citenamefont [1]{#1}%
\providecommand \href@noop [0]{\@secondoftwo}%
\providecommand \href [0]{\begingroup \@sanitize@url \@href}%
\providecommand \@href[1]{\@@startlink{#1}\@@href}%
\providecommand \@@href[1]{\endgroup#1\@@endlink}%
\providecommand \@sanitize@url [0]{\catcode `\\12\catcode `\$12\catcode
  `\&12\catcode `\#12\catcode `\^12\catcode `\_12\catcode `\%12\relax}%
\providecommand \@@startlink[1]{}%
\providecommand \@@endlink[0]{}%
\providecommand \url  [0]{\begingroup\@sanitize@url \@url }%
\providecommand \@url [1]{\endgroup\@href {#1}{\urlprefix }}%
\providecommand \urlprefix  [0]{URL }%
\providecommand \Eprint [0]{\href }%
\providecommand \doibase [0]{http://dx.doi.org/}%
\providecommand \selectlanguage [0]{\@gobble}%
\providecommand \bibinfo  [0]{\@secondoftwo}%
\providecommand \bibfield  [0]{\@secondoftwo}%
\providecommand \translation [1]{[#1]}%
\providecommand \BibitemOpen [0]{}%
\providecommand \bibitemStop [0]{}%
\providecommand \bibitemNoStop [0]{.\EOS\space}%
\providecommand \EOS [0]{\spacefactor3000\relax}%
\providecommand \BibitemShut  [1]{\csname bibitem#1\endcsname}%
\let\auto@bib@innerbib\@empty
\bibitem [{\citenamefont {Bhattacharya}\ and\ \citenamefont
  {Higgins}(1993)}]{Bhattacharya93}%
  \BibitemOpen
  \bibfield  {author} {\bibinfo {author} {\bibfnamefont {S.}~\bibnamefont
  {Bhattacharya}}\ and\ \bibinfo {author} {\bibfnamefont {M.~J.}\ \bibnamefont
  {Higgins}},\ }\bibfield  {title} {\enquote {\bibinfo {title} {Dynamics of a
  disordered flux line lattice},}\ }\href {\doibase
  10.1103/PhysRevLett.70.2617} {\bibfield  {journal} {\bibinfo  {journal}
  {Phys. Rev. Lett.}\ }\textbf {\bibinfo {volume} {70}},\ \bibinfo {pages}
  {2617--2620} (\bibinfo {year} {1993})}\BibitemShut {NoStop}%
\bibitem [{\citenamefont {Reichhardt}\ and\ \citenamefont
  {Reichhardt}(2017)}]{Reichhardt17}%
  \BibitemOpen
  \bibfield  {author} {\bibinfo {author} {\bibfnamefont {C.}~\bibnamefont
  {Reichhardt}}\ and\ \bibinfo {author} {\bibfnamefont {C.~J.~Olson}\
  \bibnamefont {Reichhardt}},\ }\bibfield  {title} {\enquote {\bibinfo {title}
  {Depinning and nonequilibrium dynamic phases of particle assemblies driven
  over random and ordered substrates: a review},}\ }\href {\doibase
  10.1088/1361-6633/80/2/026501} {\bibfield  {journal} {\bibinfo  {journal}
  {Rep. Prog. Phys.}\ }\textbf {\bibinfo {volume} {80}},\ \bibinfo {pages}
  {026501} (\bibinfo {year} {2017})}\BibitemShut {NoStop}%
\bibitem [{\citenamefont {Pertsinidis}\ and\ \citenamefont
  {Ling}(2008)}]{Pertsinidis08}%
  \BibitemOpen
  \bibfield  {author} {\bibinfo {author} {\bibfnamefont {A.}~\bibnamefont
  {Pertsinidis}}\ and\ \bibinfo {author} {\bibfnamefont {X.~S.}\ \bibnamefont
  {Ling}},\ }\bibfield  {title} {\enquote {\bibinfo {title} {Statics and
  dynamics of {2D} colloidal crystals in a random pinning potential},}\ }\href
  {\doibase 10.1103/PhysRevLett.100.028303} {\bibfield  {journal} {\bibinfo
  {journal} {Phys. Rev. Lett.}\ }\textbf {\bibinfo {volume} {100}},\ \bibinfo
  {pages} {028303} (\bibinfo {year} {2008})}\BibitemShut {NoStop}%
\bibitem [{\citenamefont {Jiang}\ \emph {et~al.}(2017)\citenamefont {Jiang},
  \citenamefont {Zhang}, \citenamefont {Yu}, \citenamefont {Zhang},
  \citenamefont {Wang}, \citenamefont {Jungfleisch}, \citenamefont {Pearson},
  \citenamefont {Cheng}, \citenamefont {Heinonen}, \citenamefont {Wang},
  \citenamefont {Zhou}, \citenamefont {Hoffmann},\ and\ \citenamefont
  {te~Velthuis}}]{Jiang17}%
  \BibitemOpen
  \bibfield  {author} {\bibinfo {author} {\bibfnamefont {W.}~\bibnamefont
  {Jiang}}, \bibinfo {author} {\bibfnamefont {X.}~\bibnamefont {Zhang}},
  \bibinfo {author} {\bibfnamefont {G.}~\bibnamefont {Yu}}, \bibinfo {author}
  {\bibfnamefont {W.}~\bibnamefont {Zhang}}, \bibinfo {author} {\bibfnamefont
  {X.}~\bibnamefont {Wang}}, \bibinfo {author} {\bibfnamefont {M.~B.}\
  \bibnamefont {Jungfleisch}}, \bibinfo {author} {\bibfnamefont {J.~E.}\
  \bibnamefont {Pearson}}, \bibinfo {author} {\bibfnamefont {X.}~\bibnamefont
  {Cheng}}, \bibinfo {author} {\bibfnamefont {O.}~\bibnamefont {Heinonen}},
  \bibinfo {author} {\bibfnamefont {K.~L.}\ \bibnamefont {Wang}}, \bibinfo
  {author} {\bibfnamefont {Y.}~\bibnamefont {Zhou}}, \bibinfo {author}
  {\bibfnamefont {A.}~\bibnamefont {Hoffmann}}, \ and\ \bibinfo {author}
  {\bibfnamefont {S.~G.~E.}\ \bibnamefont {te~Velthuis}},\ }\bibfield  {title}
  {\enquote {\bibinfo {title} {Direct observation of the skyrmion {H}all
  effect},}\ }\href {\doibase 10.1038/NPHYS3883} {\bibfield  {journal}
  {\bibinfo  {journal} {Nature Phys.}\ }\textbf {\bibinfo {volume} {13}},\
  \bibinfo {pages} {162--169} (\bibinfo {year} {2017})}\BibitemShut {NoStop}%
\bibitem [{\citenamefont {Reichhardt}\ \emph {et~al.}(2022)\citenamefont
  {Reichhardt}, \citenamefont {Reichhardt},\ and\ \citenamefont {Milo{\v
  s}evi{\' c}}}]{Reichhardt22a}%
  \BibitemOpen
  \bibfield  {author} {\bibinfo {author} {\bibfnamefont {C.}~\bibnamefont
  {Reichhardt}}, \bibinfo {author} {\bibfnamefont {C.~J.~O.}\ \bibnamefont
  {Reichhardt}}, \ and\ \bibinfo {author} {\bibfnamefont {M.}~\bibnamefont
  {Milo{\v s}evi{\' c}}},\ }\bibfield  {title} {\enquote {\bibinfo {title}
  {Statics and dynamics of skyrmions interacting with disorder and
  nanostructures},}\ }\href {\doibase 10.1103/RevModPhys.94.035005} {\bibfield
  {journal} {\bibinfo  {journal} {Rev. Mod. Phys.}\ }\textbf {\bibinfo {volume}
  {94}},\ \bibinfo {pages} {035005} (\bibinfo {year} {2022})}\BibitemShut
  {NoStop}%
\bibitem [{\citenamefont {Le~Blay}\ \emph {et~al.}(2020)\citenamefont
  {Le~Blay}, \citenamefont {Adda-Bedia},\ and\ \citenamefont
  {Bartolo}}]{LeBlay20}%
  \BibitemOpen
  \bibfield  {author} {\bibinfo {author} {\bibfnamefont {M.}~\bibnamefont
  {Le~Blay}}, \bibinfo {author} {\bibfnamefont {M.}~\bibnamefont {Adda-Bedia}},
  \ and\ \bibinfo {author} {\bibfnamefont {D.}~\bibnamefont {Bartolo}},\
  }\bibfield  {title} {\enquote {\bibinfo {title} {Emergence of scale-free
  smectic rivers and critical depinning in emulsions driven through
  disorder},}\ }\href {\doibase 10.1073/pnas.2000681117} {\bibfield  {journal}
  {\bibinfo  {journal} {Proc. Natl. Acad. Sci. (USA)}\ }\textbf {\bibinfo
  {volume} {117}},\ \bibinfo {pages} {13914--13920} (\bibinfo {year}
  {2020})}\BibitemShut {NoStop}%
\bibitem [{\citenamefont {Morin}\ \emph {et~al.}(2017)\citenamefont {Morin},
  \citenamefont {Desreumaux}, \citenamefont {Caussin},\ and\ \citenamefont
  {Bartolo}}]{Morin17}%
  \BibitemOpen
  \bibfield  {author} {\bibinfo {author} {\bibfnamefont {A.}~\bibnamefont
  {Morin}}, \bibinfo {author} {\bibfnamefont {N.}~\bibnamefont {Desreumaux}},
  \bibinfo {author} {\bibfnamefont {J.-B.}\ \bibnamefont {Caussin}}, \ and\
  \bibinfo {author} {\bibfnamefont {D.}~\bibnamefont {Bartolo}},\ }\bibfield
  {title} {\enquote {\bibinfo {title} {Distortion and destruction of colloidal
  flocks in disordered environments},}\ }\href {\doibase 10.1038/nphys3903}
  {\bibfield  {journal} {\bibinfo  {journal} {Nature Phys.}\ }\textbf {\bibinfo
  {volume} {13}},\ \bibinfo {pages} {63--67} (\bibinfo {year}
  {2017})}\BibitemShut {NoStop}%
\bibitem [{\citenamefont {S\'andor}\ \emph {et~al.}(2017)\citenamefont
  {S\'andor}, \citenamefont {Lib\'al}, \citenamefont {Reichhardt},\ and\
  \citenamefont {Olson~Reichhardt}}]{Sandor17a}%
  \BibitemOpen
  \bibfield  {author} {\bibinfo {author} {\bibfnamefont {Cs.}\ \bibnamefont
  {S\'andor}}, \bibinfo {author} {\bibfnamefont {A.}~\bibnamefont {Lib\'al}},
  \bibinfo {author} {\bibfnamefont {C.}~\bibnamefont {Reichhardt}}, \ and\
  \bibinfo {author} {\bibfnamefont {C.~J.}\ \bibnamefont {Olson~Reichhardt}},\
  }\bibfield  {title} {\enquote {\bibinfo {title} {Dynamic phases of active
  matter systems with quenched disorder},}\ }\href {\doibase
  10.1103/PhysRevE.95.032606} {\bibfield  {journal} {\bibinfo  {journal} {Phys.
  Rev. E}\ }\textbf {\bibinfo {volume} {95}},\ \bibinfo {pages} {032606}
  (\bibinfo {year} {2017})}\BibitemShut {NoStop}%
\bibitem [{\citenamefont {Olson~Reichhardt}\ \emph {et~al.}(2012)\citenamefont
  {Olson~Reichhardt}, \citenamefont {Groopman}, \citenamefont {Nussinov},\ and\
  \citenamefont {Reichhardt}}]{Reichhardt12}%
  \BibitemOpen
  \bibfield  {author} {\bibinfo {author} {\bibfnamefont {C.~J.}\ \bibnamefont
  {Olson~Reichhardt}}, \bibinfo {author} {\bibfnamefont {E.}~\bibnamefont
  {Groopman}}, \bibinfo {author} {\bibfnamefont {Z.}~\bibnamefont {Nussinov}},
  \ and\ \bibinfo {author} {\bibfnamefont {C.}~\bibnamefont {Reichhardt}},\
  }\bibfield  {title} {\enquote {\bibinfo {title} {Jamming in systems with
  quenched disorder},}\ }\href {\doibase 10.1103/PhysRevE.86.061301} {\bibfield
   {journal} {\bibinfo  {journal} {Phys. Rev. E}\ }\textbf {\bibinfo {volume}
  {86}},\ \bibinfo {pages} {061301} (\bibinfo {year} {2012})}\BibitemShut
  {NoStop}%
\bibitem [{\citenamefont {Zuriguel}\ \emph {et~al.}(2015)\citenamefont
  {Zuriguel}, \citenamefont {Parisi}, \citenamefont {Hidalgo}, \citenamefont
  {Lozano}, \citenamefont {Janda}, \citenamefont {Gago}, \citenamefont
  {Peralta}, \citenamefont {Ferrer}, \citenamefont {Pugnaloni}, \citenamefont
  {Cl{\' e}ment}, \citenamefont {Maza}, \citenamefont {Pagonabarraga},\ and\
  \citenamefont {Garcimart{\' \i}n}}]{Zuriguel15}%
  \BibitemOpen
  \bibfield  {author} {\bibinfo {author} {\bibfnamefont {I.}~\bibnamefont
  {Zuriguel}}, \bibinfo {author} {\bibfnamefont {D.~R.}\ \bibnamefont
  {Parisi}}, \bibinfo {author} {\bibfnamefont {R.~C.}\ \bibnamefont {Hidalgo}},
  \bibinfo {author} {\bibfnamefont {C.}~\bibnamefont {Lozano}}, \bibinfo
  {author} {\bibfnamefont {A.}~\bibnamefont {Janda}}, \bibinfo {author}
  {\bibfnamefont {P.~A.}\ \bibnamefont {Gago}}, \bibinfo {author}
  {\bibfnamefont {J.~P.}\ \bibnamefont {Peralta}}, \bibinfo {author}
  {\bibfnamefont {L.~M.}\ \bibnamefont {Ferrer}}, \bibinfo {author}
  {\bibfnamefont {L.~A.}\ \bibnamefont {Pugnaloni}}, \bibinfo {author}
  {\bibfnamefont {E.}~\bibnamefont {Cl{\' e}ment}}, \bibinfo {author}
  {\bibfnamefont {D.}~\bibnamefont {Maza}}, \bibinfo {author} {\bibfnamefont
  {I.}~\bibnamefont {Pagonabarraga}}, \ and\ \bibinfo {author} {\bibfnamefont
  {A.}~\bibnamefont {Garcimart{\' \i}n}},\ }\bibfield  {title} {\enquote
  {\bibinfo {title} {Clogging transition of many-particle systems flowing
  through bottlenecks},}\ }\href {\doibase 10.1038/srep07324} {\bibfield
  {journal} {\bibinfo  {journal} {Sci. Rep.}\ }\textbf {\bibinfo {volume}
  {4}},\ \bibinfo {pages} {7324} (\bibinfo {year} {2015})}\BibitemShut
  {NoStop}%
\bibitem [{\citenamefont {Nguyen}\ \emph {et~al.}(2017)\citenamefont {Nguyen},
  \citenamefont {Reichhardt},\ and\ \citenamefont {Reichhardt}}]{Nguyen17}%
  \BibitemOpen
  \bibfield  {author} {\bibinfo {author} {\bibfnamefont {H.~T.}\ \bibnamefont
  {Nguyen}}, \bibinfo {author} {\bibfnamefont {C.}~\bibnamefont {Reichhardt}},
  \ and\ \bibinfo {author} {\bibfnamefont {C.~J.~Olson}\ \bibnamefont
  {Reichhardt}},\ }\bibfield  {title} {\enquote {\bibinfo {title} {Clogging and
  jamming transitions in periodic obstacle arrays},}\ }\href {\doibase
  10.1103/PhysRevE.95.030902} {\bibfield  {journal} {\bibinfo  {journal} {Phys.
  Rev. E}\ }\textbf {\bibinfo {volume} {95}},\ \bibinfo {pages} {030902}
  (\bibinfo {year} {2017})}\BibitemShut {NoStop}%
\bibitem [{\citenamefont {Gerber}\ \emph {et~al.}(2018)\citenamefont {Gerber},
  \citenamefont {Rodts}, \citenamefont {Aimedieu}, \citenamefont {Faure},\ and\
  \citenamefont {Coussot}}]{Gerber18}%
  \BibitemOpen
  \bibfield  {author} {\bibinfo {author} {\bibfnamefont {G.}~\bibnamefont
  {Gerber}}, \bibinfo {author} {\bibfnamefont {S.}~\bibnamefont {Rodts}},
  \bibinfo {author} {\bibfnamefont {P.}~\bibnamefont {Aimedieu}}, \bibinfo
  {author} {\bibfnamefont {P.}~\bibnamefont {Faure}}, \ and\ \bibinfo {author}
  {\bibfnamefont {P.}~\bibnamefont {Coussot}},\ }\bibfield  {title} {\enquote
  {\bibinfo {title} {Particle-size-exclusion clogging regimes in porous
  media},}\ }\href {\doibase 10.1103/PhysRevLett.120.148001} {\bibfield
  {journal} {\bibinfo  {journal} {Phys. Rev. Lett.}\ }\textbf {\bibinfo
  {volume} {120}},\ \bibinfo {pages} {148001} (\bibinfo {year}
  {2018})}\BibitemShut {NoStop}%
\bibitem [{\citenamefont {Stoop}\ and\ \citenamefont {Tierno}(2018)}]{Stoop18}%
  \BibitemOpen
  \bibfield  {author} {\bibinfo {author} {\bibfnamefont {R.~L.}\ \bibnamefont
  {Stoop}}\ and\ \bibinfo {author} {\bibfnamefont {P.}~\bibnamefont {Tierno}},\
  }\bibfield  {title} {\enquote {\bibinfo {title} {Clogging and jamming of
  colloidal monolayers driven across disordered landscapes},}\ }\href {\doibase
  10.1038/s42005-018-0068-6} {\bibfield  {journal} {\bibinfo  {journal}
  {Commun. Phys.}\ }\textbf {\bibinfo {volume} {1}},\ \bibinfo {pages} {68}
  (\bibinfo {year} {2018})}\BibitemShut {NoStop}%
\bibitem [{\citenamefont {Reichhardt}\ \emph {et~al.}(2023)\citenamefont
  {Reichhardt}, \citenamefont {Regev}, \citenamefont {Dahmen}, \citenamefont
  {Okuma},\ and\ \citenamefont {Reichhardt}}]{Reichhardt23}%
  \BibitemOpen
  \bibfield  {author} {\bibinfo {author} {\bibfnamefont {C.}~\bibnamefont
  {Reichhardt}}, \bibinfo {author} {\bibfnamefont {Ido}\ \bibnamefont {Regev}},
  \bibinfo {author} {\bibfnamefont {K.}~\bibnamefont {Dahmen}}, \bibinfo
  {author} {\bibfnamefont {S.}~\bibnamefont {Okuma}}, \ and\ \bibinfo {author}
  {\bibfnamefont {C.~J.~O.}\ \bibnamefont {Reichhardt}},\ }\bibfield  {title}
  {\enquote {\bibinfo {title} {Reversible to irreversible transitions in
  periodic driven many-body systems and future directions for classical and
  quantum systems},}\ }\href {\doibase 10.1103/PhysRevResearch.5.021001}
  {\bibfield  {journal} {\bibinfo  {journal} {Phys. Rev. Res.}\ }\textbf
  {\bibinfo {volume} {5}},\ \bibinfo {pages} {021001} (\bibinfo {year}
  {2023})}\BibitemShut {NoStop}%
\bibitem [{\citenamefont {Pine}\ \emph {et~al.}(2005)\citenamefont {Pine},
  \citenamefont {Gollub}, \citenamefont {Brady},\ and\ \citenamefont
  {Leshansky}}]{Pine05}%
  \BibitemOpen
  \bibfield  {author} {\bibinfo {author} {\bibfnamefont {D.~J.}\ \bibnamefont
  {Pine}}, \bibinfo {author} {\bibfnamefont {J.~P.}\ \bibnamefont {Gollub}},
  \bibinfo {author} {\bibfnamefont {J.~F.}\ \bibnamefont {Brady}}, \ and\
  \bibinfo {author} {\bibfnamefont {A.~M.}\ \bibnamefont {Leshansky}},\
  }\bibfield  {title} {\enquote {\bibinfo {title} {Chaos and threshold for
  irreversibility in sheared suspensions},}\ }\href {\doibase
  10.1038/nature04380} {\bibfield  {journal} {\bibinfo  {journal} {Nature
  (London)}\ }\textbf {\bibinfo {volume} {438}},\ \bibinfo {pages} {997--1000}
  (\bibinfo {year} {2005})}\BibitemShut {NoStop}%
\bibitem [{\citenamefont {Corte}\ \emph {et~al.}(2008)\citenamefont {Corte},
  \citenamefont {Chaikin}, \citenamefont {Gollub},\ and\ \citenamefont
  {Pine}}]{Corte08}%
  \BibitemOpen
  \bibfield  {author} {\bibinfo {author} {\bibfnamefont {L.}~\bibnamefont
  {Corte}}, \bibinfo {author} {\bibfnamefont {P.~M.}\ \bibnamefont {Chaikin}},
  \bibinfo {author} {\bibfnamefont {J.~P.}\ \bibnamefont {Gollub}}, \ and\
  \bibinfo {author} {\bibfnamefont {D.~J.}\ \bibnamefont {Pine}},\ }\bibfield
  {title} {\enquote {\bibinfo {title} {Random organization in periodically
  driven systems},}\ }\href {\doibase 10.1038/nphys891} {\bibfield  {journal}
  {\bibinfo  {journal} {Nature Phys.}\ }\textbf {\bibinfo {volume} {4}},\
  \bibinfo {pages} {420--424} (\bibinfo {year} {2008})}\BibitemShut {NoStop}%
\bibitem [{\citenamefont {Hexner}\ and\ \citenamefont
  {Levine}(2015)}]{Hexner15}%
  \BibitemOpen
  \bibfield  {author} {\bibinfo {author} {\bibfnamefont {D.}~\bibnamefont
  {Hexner}}\ and\ \bibinfo {author} {\bibfnamefont {D.}~\bibnamefont
  {Levine}},\ }\bibfield  {title} {\enquote {\bibinfo {title} {Hyperuniformity
  of critical absorbing states},}\ }\href {\doibase
  10.1103/PhysRevLett.114.110602} {\bibfield  {journal} {\bibinfo  {journal}
  {Phys. Rev. Lett.}\ }\textbf {\bibinfo {volume} {114}},\ \bibinfo {pages}
  {110602} (\bibinfo {year} {2015})}\BibitemShut {NoStop}%
\bibitem [{\citenamefont {Tjhung}\ and\ \citenamefont
  {Berthier}(2015)}]{Tjhung15}%
  \BibitemOpen
  \bibfield  {author} {\bibinfo {author} {\bibfnamefont {E.}~\bibnamefont
  {Tjhung}}\ and\ \bibinfo {author} {\bibfnamefont {L.}~\bibnamefont
  {Berthier}},\ }\bibfield  {title} {\enquote {\bibinfo {title} {Hyperuniform
  density fluctuations and diverging dynamic correlations in periodically
  driven colloidal suspensions},}\ }\href {\doibase
  10.1103/PhysRevLett.114.148301} {\bibfield  {journal} {\bibinfo  {journal}
  {Phys. Rev. Lett.}\ }\textbf {\bibinfo {volume} {114}},\ \bibinfo {pages}
  {148301} (\bibinfo {year} {2015})}\BibitemShut {NoStop}%
\bibitem [{\citenamefont {Weijs}\ \emph {et~al.}(2015)\citenamefont {Weijs},
  \citenamefont {Jeanneret}, \citenamefont {Dreyfus},\ and\ \citenamefont
  {Bartolo}}]{Weijs15}%
  \BibitemOpen
  \bibfield  {author} {\bibinfo {author} {\bibfnamefont {J.~H.}\ \bibnamefont
  {Weijs}}, \bibinfo {author} {\bibfnamefont {R.}~\bibnamefont {Jeanneret}},
  \bibinfo {author} {\bibfnamefont {R.}~\bibnamefont {Dreyfus}}, \ and\
  \bibinfo {author} {\bibfnamefont {D.}~\bibnamefont {Bartolo}},\ }\bibfield
  {title} {\enquote {\bibinfo {title} {Emergent hyperuniformity in periodically
  driven emulsions},}\ }\href {\doibase 10.1103/PhysRevLett.115.108301}
  {\bibfield  {journal} {\bibinfo  {journal} {Phys. Rev. Lett.}\ }\textbf
  {\bibinfo {volume} {115}},\ \bibinfo {pages} {108301} (\bibinfo {year}
  {2015})}\BibitemShut {NoStop}%
\bibitem [{\citenamefont {Reichhardt}\ and\ \citenamefont
  {Reichhardt}(2019)}]{Reichhardt19}%
  \BibitemOpen
  \bibfield  {author} {\bibinfo {author} {\bibfnamefont {C.}~\bibnamefont
  {Reichhardt}}\ and\ \bibinfo {author} {\bibfnamefont {C.~J.~O.}\ \bibnamefont
  {Reichhardt}},\ }\bibfield  {title} {\enquote {\bibinfo {title}
  {Reversibility, pattern formation, and edge transport in active chiral and
  passive disk mixtures},}\ }\href {\doibase 10.1063/1.5085209} {\bibfield
  {journal} {\bibinfo  {journal} {J. Chem. Phys.}\ }\textbf {\bibinfo {volume}
  {150}},\ \bibinfo {pages} {064905} (\bibinfo {year} {2019})}\BibitemShut
  {NoStop}%
\bibitem [{\citenamefont {Lei}\ and\ \citenamefont {Ni}(2019)}]{Lei19a}%
  \BibitemOpen
  \bibfield  {author} {\bibinfo {author} {\bibfnamefont {Q.-L.}\ \bibnamefont
  {Lei}}\ and\ \bibinfo {author} {\bibfnamefont {R.}~\bibnamefont {Ni}},\
  }\bibfield  {title} {\enquote {\bibinfo {title} {Hydrodynamics of
  random-organizing hyperuniform fluids},}\ }\href {\doibase
  10.1073/pnas.1911596116} {\bibfield  {journal} {\bibinfo  {journal} {Proc.
  Natl. Acad. Sci. (USA)}\ }\textbf {\bibinfo {volume} {116}},\ \bibinfo
  {pages} {22983} (\bibinfo {year} {2019})}\BibitemShut {NoStop}%
\bibitem [{\citenamefont {Paulsen}\ \emph {et~al.}(2014)\citenamefont
  {Paulsen}, \citenamefont {Keim},\ and\ \citenamefont {Nagel}}]{Paulsen14}%
  \BibitemOpen
  \bibfield  {author} {\bibinfo {author} {\bibfnamefont {J.~D.}\ \bibnamefont
  {Paulsen}}, \bibinfo {author} {\bibfnamefont {N.~C.}\ \bibnamefont {Keim}}, \
  and\ \bibinfo {author} {\bibfnamefont {S.~R.}\ \bibnamefont {Nagel}},\
  }\bibfield  {title} {\enquote {\bibinfo {title} {Multiple transient memories
  in experiments on sheared non-{B}rownian suspensions},}\ }\href {\doibase
  10.1103/PhysRevLett.113.068301} {\bibfield  {journal} {\bibinfo  {journal}
  {Phys. Rev. Lett.}\ }\textbf {\bibinfo {volume} {113}},\ \bibinfo {pages}
  {068301} (\bibinfo {year} {2014})}\BibitemShut {NoStop}%
\bibitem [{\citenamefont {Keim}\ \emph {et~al.}(2019)\citenamefont {Keim},
  \citenamefont {Paulsen}, \citenamefont {Zeravcic}, \citenamefont {Sastry},\
  and\ \citenamefont {Nagel}}]{Keim19}%
  \BibitemOpen
  \bibfield  {author} {\bibinfo {author} {\bibfnamefont {N.~C.}\ \bibnamefont
  {Keim}}, \bibinfo {author} {\bibfnamefont {J.~D.}\ \bibnamefont {Paulsen}},
  \bibinfo {author} {\bibfnamefont {Z.}~\bibnamefont {Zeravcic}}, \bibinfo
  {author} {\bibfnamefont {S.}~\bibnamefont {Sastry}}, \ and\ \bibinfo {author}
  {\bibfnamefont {S.~R.}\ \bibnamefont {Nagel}},\ }\bibfield  {title} {\enquote
  {\bibinfo {title} {Memory formation in matter},}\ }\href {\doibase
  10.1103/RevModPhys.91.035002} {\bibfield  {journal} {\bibinfo  {journal}
  {Rev. Mod. Phys.}\ }\textbf {\bibinfo {volume} {91}},\ \bibinfo {pages}
  {035002} (\bibinfo {year} {2019})}\BibitemShut {NoStop}%
\bibitem [{\citenamefont {Schreck}\ \emph {et~al.}(2013)\citenamefont
  {Schreck}, \citenamefont {Hoy}, \citenamefont {Shattuck},\ and\ \citenamefont
  {O'Hern}}]{Schreck13}%
  \BibitemOpen
  \bibfield  {author} {\bibinfo {author} {\bibfnamefont {C.~F.}\ \bibnamefont
  {Schreck}}, \bibinfo {author} {\bibfnamefont {R.~S.}\ \bibnamefont {Hoy}},
  \bibinfo {author} {\bibfnamefont {M.~D.}\ \bibnamefont {Shattuck}}, \ and\
  \bibinfo {author} {\bibfnamefont {C.~S.}\ \bibnamefont {O'Hern}},\ }\bibfield
   {title} {\enquote {\bibinfo {title} {Particle-scale reversibility in
  athermal particulate media below jamming},}\ }\href {\doibase
  10.1103/PhysRevE.88.052205} {\bibfield  {journal} {\bibinfo  {journal} {Phys.
  Rev. E}\ }\textbf {\bibinfo {volume} {88}},\ \bibinfo {pages} {052205}
  (\bibinfo {year} {2013})}\BibitemShut {NoStop}%
\bibitem [{\citenamefont {Royer}\ and\ \citenamefont
  {Chaikin}(2015)}]{Royer15}%
  \BibitemOpen
  \bibfield  {author} {\bibinfo {author} {\bibfnamefont {J.~R.}\ \bibnamefont
  {Royer}}\ and\ \bibinfo {author} {\bibfnamefont {P.~M.}\ \bibnamefont
  {Chaikin}},\ }\bibfield  {title} {\enquote {\bibinfo {title} {Precisely
  cyclic sand: Self-organization of periodically sheared frictional grains},}\
  }\href {\doibase 10.1073/pnas.1413468112} {\bibfield  {journal} {\bibinfo
  {journal} {Proc. Natl. Acad. Sci. (USA)}\ }\textbf {\bibinfo {volume}
  {112}},\ \bibinfo {pages} {49--53} (\bibinfo {year} {2015})}\BibitemShut
  {NoStop}%
\bibitem [{\citenamefont {Regev}\ \emph {et~al.}(2013)\citenamefont {Regev},
  \citenamefont {Lookman},\ and\ \citenamefont {Reichhardt}}]{Regev13}%
  \BibitemOpen
  \bibfield  {author} {\bibinfo {author} {\bibfnamefont {I.}~\bibnamefont
  {Regev}}, \bibinfo {author} {\bibfnamefont {T.}~\bibnamefont {Lookman}}, \
  and\ \bibinfo {author} {\bibfnamefont {C.}~\bibnamefont {Reichhardt}},\
  }\bibfield  {title} {\enquote {\bibinfo {title} {Onset of irreversibility and
  chaos in amorphous solids under periodic shear},}\ }\href {\doibase
  10.1103/PhysRevE.88.062401} {\bibfield  {journal} {\bibinfo  {journal} {Phys.
  Rev. E}\ }\textbf {\bibinfo {volume} {88}},\ \bibinfo {pages} {062401}
  (\bibinfo {year} {2013})}\BibitemShut {NoStop}%
\bibitem [{\citenamefont {Fiocco}\ \emph {et~al.}(2013)\citenamefont {Fiocco},
  \citenamefont {Foffi},\ and\ \citenamefont {Sastry}}]{Fiocco13}%
  \BibitemOpen
  \bibfield  {author} {\bibinfo {author} {\bibfnamefont {D.}~\bibnamefont
  {Fiocco}}, \bibinfo {author} {\bibfnamefont {G.}~\bibnamefont {Foffi}}, \
  and\ \bibinfo {author} {\bibfnamefont {S.}~\bibnamefont {Sastry}},\
  }\bibfield  {title} {\enquote {\bibinfo {title} {Oscillatory athermal
  quasistatic deformation of a model glass},}\ }\href {\doibase
  10.1103/PhysRevE.88.020301} {\bibfield  {journal} {\bibinfo  {journal} {Phys.
  Rev. E}\ }\textbf {\bibinfo {volume} {88}},\ \bibinfo {pages} {020301}
  (\bibinfo {year} {2013})}\BibitemShut {NoStop}%
\bibitem [{\citenamefont {Fiocco}\ \emph {et~al.}(2014)\citenamefont {Fiocco},
  \citenamefont {Foffi},\ and\ \citenamefont {Sastry}}]{Fiocco14}%
  \BibitemOpen
  \bibfield  {author} {\bibinfo {author} {\bibfnamefont {D.}~\bibnamefont
  {Fiocco}}, \bibinfo {author} {\bibfnamefont {G.}~\bibnamefont {Foffi}}, \
  and\ \bibinfo {author} {\bibfnamefont {S.}~\bibnamefont {Sastry}},\
  }\bibfield  {title} {\enquote {\bibinfo {title} {Encoding of memory in
  sheared amorphous solids},}\ }\href {\doibase 10.1103/PhysRevLett.112.025702}
  {\bibfield  {journal} {\bibinfo  {journal} {Phys. Rev. Lett.}\ }\textbf
  {\bibinfo {volume} {112}},\ \bibinfo {pages} {025702} (\bibinfo {year}
  {2014})}\BibitemShut {NoStop}%
\bibitem [{\citenamefont {Keim}\ and\ \citenamefont {Arratia}(2014)}]{Keim14}%
  \BibitemOpen
  \bibfield  {author} {\bibinfo {author} {\bibfnamefont {N.~C.}\ \bibnamefont
  {Keim}}\ and\ \bibinfo {author} {\bibfnamefont {P.~E.}\ \bibnamefont
  {Arratia}},\ }\bibfield  {title} {\enquote {\bibinfo {title} {Mechanical and
  microscopic properties of the reversible plastic regime in a {2D} jammed
  material},}\ }\href {\doibase 10.1103/PhysRevLett.112.028302} {\bibfield
  {journal} {\bibinfo  {journal} {Phys. Rev. Lett.}\ }\textbf {\bibinfo
  {volume} {112}},\ \bibinfo {pages} {028302} (\bibinfo {year}
  {2014})}\BibitemShut {NoStop}%
\bibitem [{\citenamefont {Priezjev}(2016)}]{Priezjev16}%
  \BibitemOpen
  \bibfield  {author} {\bibinfo {author} {\bibfnamefont {N.~V.}\ \bibnamefont
  {Priezjev}},\ }\bibfield  {title} {\enquote {\bibinfo {title} {Reversible
  plastic events during oscillatory deformation of amorphous solids},}\ }\href
  {\doibase 10.1103/PhysRevE.93.013001} {\bibfield  {journal} {\bibinfo
  {journal} {Phys. Rev. E}\ }\textbf {\bibinfo {volume} {93}},\ \bibinfo
  {pages} {013001} (\bibinfo {year} {2016})}\BibitemShut {NoStop}%
\bibitem [{\citenamefont {Priezjev}(2017)}]{Priezjev17}%
  \BibitemOpen
  \bibfield  {author} {\bibinfo {author} {\bibfnamefont {N.~V.}\ \bibnamefont
  {Priezjev}},\ }\bibfield  {title} {\enquote {\bibinfo {title} {Collective
  nonaffine displacements in amorphous materials during large-amplitude
  oscillatory shear},}\ }\href {\doibase 10.1103/PhysRevE.95.023002} {\bibfield
   {journal} {\bibinfo  {journal} {Phys. Rev. E}\ }\textbf {\bibinfo {volume}
  {95}},\ \bibinfo {pages} {023002} (\bibinfo {year} {2017})}\BibitemShut
  {NoStop}%
\bibitem [{\citenamefont {Lavrentovich}\ \emph {et~al.}(2017)\citenamefont
  {Lavrentovich}, \citenamefont {Liu},\ and\ \citenamefont
  {Nagel}}]{Lavrentovich17}%
  \BibitemOpen
  \bibfield  {author} {\bibinfo {author} {\bibfnamefont {M.~O.}\ \bibnamefont
  {Lavrentovich}}, \bibinfo {author} {\bibfnamefont {A.~J.}\ \bibnamefont
  {Liu}}, \ and\ \bibinfo {author} {\bibfnamefont {S.~R.}\ \bibnamefont
  {Nagel}},\ }\bibfield  {title} {\enquote {\bibinfo {title} {Period
  proliferation in periodic states in cyclically sheared jammed solids},}\
  }\href {\doibase 10.1103/PhysRevE.96.020101} {\bibfield  {journal} {\bibinfo
  {journal} {Phys. Rev. E}\ }\textbf {\bibinfo {volume} {96}},\ \bibinfo
  {pages} {020101} (\bibinfo {year} {2017})}\BibitemShut {NoStop}%
\bibitem [{\citenamefont {Khirallah}\ \emph {et~al.}(2021)\citenamefont
  {Khirallah}, \citenamefont {Tyukodi}, \citenamefont {Vandembroucq},\ and\
  \citenamefont {Maloney}}]{Khirallah21}%
  \BibitemOpen
  \bibfield  {author} {\bibinfo {author} {\bibfnamefont {K.}~\bibnamefont
  {Khirallah}}, \bibinfo {author} {\bibfnamefont {B.}~\bibnamefont {Tyukodi}},
  \bibinfo {author} {\bibfnamefont {D.}~\bibnamefont {Vandembroucq}}, \ and\
  \bibinfo {author} {\bibfnamefont {C.~E.}\ \bibnamefont {Maloney}},\
  }\bibfield  {title} {\enquote {\bibinfo {title} {Yielding in an integer
  automaton model for amorphous solids under cyclic shear},}\ }\href {\doibase
  10.1103/PhysRevLett.126.218005} {\bibfield  {journal} {\bibinfo  {journal}
  {Phys. Rev. Lett.}\ }\textbf {\bibinfo {volume} {126}},\ \bibinfo {pages}
  {218005} (\bibinfo {year} {2021})}\BibitemShut {NoStop}%
\bibitem [{\citenamefont {Keim}\ and\ \citenamefont {Paulsen}(2021)}]{Keim21}%
  \BibitemOpen
  \bibfield  {author} {\bibinfo {author} {\bibfnamefont {N.~C.}\ \bibnamefont
  {Keim}}\ and\ \bibinfo {author} {\bibfnamefont {J.~D.}\ \bibnamefont
  {Paulsen}},\ }\bibfield  {title} {\enquote {\bibinfo {title} {Multiperiodic
  orbits from interacting soft spots in cyclically sheared amorphous solids},}\
  }\href {\doibase 10.1126/sciadv.abg7685} {\bibfield  {journal} {\bibinfo
  {journal} {Sci. Adv.}\ }\textbf {\bibinfo {volume} {7}},\ \bibinfo {pages}
  {eabg7685} (\bibinfo {year} {2021})}\BibitemShut {NoStop}%
\bibitem [{\citenamefont {Mangan}\ \emph {et~al.}(2008)\citenamefont {Mangan},
  \citenamefont {Reichhardt},\ and\ \citenamefont {Reichhardt}}]{Mangan08}%
  \BibitemOpen
  \bibfield  {author} {\bibinfo {author} {\bibfnamefont {N.}~\bibnamefont
  {Mangan}}, \bibinfo {author} {\bibfnamefont {C.}~\bibnamefont {Reichhardt}},
  \ and\ \bibinfo {author} {\bibfnamefont {C.~J.~Olson}\ \bibnamefont
  {Reichhardt}},\ }\bibfield  {title} {\enquote {\bibinfo {title} {Reversible
  to irreversible flow transition in periodically driven vortices},}\ }\href
  {\doibase 10.1103/PhysRevLett.100.187002} {\bibfield  {journal} {\bibinfo
  {journal} {Phys. Rev. Lett.}\ }\textbf {\bibinfo {volume} {100}},\ \bibinfo
  {pages} {187002} (\bibinfo {year} {2008})}\BibitemShut {NoStop}%
\bibitem [{\citenamefont {Okuma}\ \emph {et~al.}(2011)\citenamefont {Okuma},
  \citenamefont {Tsugawa},\ and\ \citenamefont {Motohashi}}]{Okuma11}%
  \BibitemOpen
  \bibfield  {author} {\bibinfo {author} {\bibfnamefont {S.}~\bibnamefont
  {Okuma}}, \bibinfo {author} {\bibfnamefont {Y.}~\bibnamefont {Tsugawa}}, \
  and\ \bibinfo {author} {\bibfnamefont {A.}~\bibnamefont {Motohashi}},\
  }\bibfield  {title} {\enquote {\bibinfo {title} {Transition from reversible
  to irreversible flow: Absorbing and depinning transitions in a sheared-vortex
  system},}\ }\href {\doibase 10.1103/PhysRevB.83.012503} {\bibfield  {journal}
  {\bibinfo  {journal} {Phys. Rev. B}\ }\textbf {\bibinfo {volume} {83}},\
  \bibinfo {pages} {012503} (\bibinfo {year} {2011})}\BibitemShut {NoStop}%
\bibitem [{\citenamefont {Pasquini}\ \emph {et~al.}(2021)\citenamefont
  {Pasquini}, \citenamefont {Berm{\'u}dez},\ and\ \citenamefont
  {Bekeris}}]{Pasquini21}%
  \BibitemOpen
  \bibfield  {author} {\bibinfo {author} {\bibfnamefont {G.}~\bibnamefont
  {Pasquini}}, \bibinfo {author} {\bibfnamefont {M.~M.}\ \bibnamefont
  {Berm{\'u}dez}}, \ and\ \bibinfo {author} {\bibfnamefont {V.}~\bibnamefont
  {Bekeris}},\ }\bibfield  {title} {\enquote {\bibinfo {title} {{AC} dynamic
  reorganization and critical phase transitions in superconducting vortex
  matter},}\ }\href {\doibase 10.1088/1361-6668/abbbc8} {\bibfield  {journal}
  {\bibinfo  {journal} {Supercond. Sci. Technol.}\ }\textbf {\bibinfo {volume}
  {34}},\ \bibinfo {pages} {013003} (\bibinfo {year} {2021})}\BibitemShut
  {NoStop}%
\bibitem [{\citenamefont {Maegochi}\ \emph {et~al.}(2019)\citenamefont
  {Maegochi}, \citenamefont {Ienaga}, \citenamefont {Kaneko},\ and\
  \citenamefont {Okuma}}]{Maegochi19}%
  \BibitemOpen
  \bibfield  {author} {\bibinfo {author} {\bibfnamefont {S.}~\bibnamefont
  {Maegochi}}, \bibinfo {author} {\bibfnamefont {K.}~\bibnamefont {Ienaga}},
  \bibinfo {author} {\bibfnamefont {S.}~\bibnamefont {Kaneko}}, \ and\ \bibinfo
  {author} {\bibfnamefont {S.}~\bibnamefont {Okuma}},\ }\bibfield  {title}
  {\enquote {\bibinfo {title} {Critical behavior near the
  reversible-irreversible transition in periodically driven vortices under
  random local shear},}\ }\href {\doibase 10.1038/s41598-019-51060-9}
  {\bibfield  {journal} {\bibinfo  {journal} {Sci. Rep.}\ }\textbf {\bibinfo
  {volume} {9}},\ \bibinfo {pages} {16447} (\bibinfo {year}
  {2019})}\BibitemShut {NoStop}%
\bibitem [{\citenamefont {Maegochi}\ \emph {et~al.}(2021)\citenamefont
  {Maegochi}, \citenamefont {Ienaga},\ and\ \citenamefont
  {Okuma}}]{Maegochi21}%
  \BibitemOpen
  \bibfield  {author} {\bibinfo {author} {\bibfnamefont {S.}~\bibnamefont
  {Maegochi}}, \bibinfo {author} {\bibfnamefont {K.}~\bibnamefont {Ienaga}}, \
  and\ \bibinfo {author} {\bibfnamefont {S.}~\bibnamefont {Okuma}},\ }\bibfield
   {title} {\enquote {\bibinfo {title} {Critical behavior of density-driven and
  shear-driven reversible-irreversible transitions in cyclically sheared
  vortices},}\ }\href {\doibase 10.1038/s41598-021-98959-w} {\bibfield
  {journal} {\bibinfo  {journal} {Sci. Rep.}\ }\textbf {\bibinfo {volume}
  {11}},\ \bibinfo {pages} {19280} (\bibinfo {year} {2021})}\BibitemShut
  {NoStop}%
\bibitem [{\citenamefont {Litzius}\ \emph {et~al.}(2017)\citenamefont
  {Litzius}, \citenamefont {Lemesh}, \citenamefont {Kr{\" u}ger}, \citenamefont
  {Bassirian}, \citenamefont {Caretta}, \citenamefont {Richter}, \citenamefont
  {B{\" u}ttner}, \citenamefont {Sato}, \citenamefont {Tretiakov},
  \citenamefont {F{\" o}rster}, \citenamefont {Reeve}, \citenamefont {Weigand},
  \citenamefont {Bykova}, \citenamefont {Stoll}, \citenamefont {Sch{\" u}tz},
  \citenamefont {Beach},\ and\ \citenamefont {Kl{\" a}ui}}]{Litzius17}%
  \BibitemOpen
  \bibfield  {author} {\bibinfo {author} {\bibfnamefont {K.}~\bibnamefont
  {Litzius}}, \bibinfo {author} {\bibfnamefont {I.}~\bibnamefont {Lemesh}},
  \bibinfo {author} {\bibfnamefont {B.}~\bibnamefont {Kr{\" u}ger}}, \bibinfo
  {author} {\bibfnamefont {P.}~\bibnamefont {Bassirian}}, \bibinfo {author}
  {\bibfnamefont {L.}~\bibnamefont {Caretta}}, \bibinfo {author} {\bibfnamefont
  {K.}~\bibnamefont {Richter}}, \bibinfo {author} {\bibfnamefont
  {F.}~\bibnamefont {B{\" u}ttner}}, \bibinfo {author} {\bibfnamefont
  {K.}~\bibnamefont {Sato}}, \bibinfo {author} {\bibfnamefont {O.~A.}\
  \bibnamefont {Tretiakov}}, \bibinfo {author} {\bibfnamefont {J.}~\bibnamefont
  {F{\" o}rster}}, \bibinfo {author} {\bibfnamefont {R.~M.}\ \bibnamefont
  {Reeve}}, \bibinfo {author} {\bibfnamefont {M.}~\bibnamefont {Weigand}},
  \bibinfo {author} {\bibfnamefont {I.}~\bibnamefont {Bykova}}, \bibinfo
  {author} {\bibfnamefont {H.}~\bibnamefont {Stoll}}, \bibinfo {author}
  {\bibfnamefont {G.}~\bibnamefont {Sch{\" u}tz}}, \bibinfo {author}
  {\bibfnamefont {G.~S.~D.}\ \bibnamefont {Beach}}, \ and\ \bibinfo {author}
  {\bibfnamefont {M.}~\bibnamefont {Kl{\" a}ui}},\ }\bibfield  {title}
  {\enquote {\bibinfo {title} {Skyrmion {H}all effect revealed by direct
  time-resolved {X}-ray microscopy},}\ }\href {\doibase 10.1038/NPHYS4000}
  {\bibfield  {journal} {\bibinfo  {journal} {Nature Phys.}\ }\textbf {\bibinfo
  {volume} {13}},\ \bibinfo {pages} {170--175} (\bibinfo {year}
  {2017})}\BibitemShut {NoStop}%
\bibitem [{\citenamefont {Brown}\ \emph {et~al.}(2018)\citenamefont {Brown},
  \citenamefont {T\"auber},\ and\ \citenamefont {Pleimling}}]{Brown18}%
  \BibitemOpen
  \bibfield  {author} {\bibinfo {author} {\bibfnamefont {B.~L.}\ \bibnamefont
  {Brown}}, \bibinfo {author} {\bibfnamefont {U.~C.}\ \bibnamefont {T\"auber}},
  \ and\ \bibinfo {author} {\bibfnamefont {M.}~\bibnamefont {Pleimling}},\
  }\bibfield  {title} {\enquote {\bibinfo {title} {Effect of the {M}agnus force
  on skyrmion relaxation dynamics},}\ }\href {\doibase
  10.1103/PhysRevB.97.020405} {\bibfield  {journal} {\bibinfo  {journal} {Phys.
  Rev. B}\ }\textbf {\bibinfo {volume} {97}},\ \bibinfo {pages} {020405}
  (\bibinfo {year} {2018})}\BibitemShut {NoStop}%
\bibitem [{\citenamefont {Reichhardt}\ and\ \citenamefont
  {Reichhardt}(2022)}]{Reichhardt22}%
  \BibitemOpen
  \bibfield  {author} {\bibinfo {author} {\bibfnamefont {C.}~\bibnamefont
  {Reichhardt}}\ and\ \bibinfo {author} {\bibfnamefont {C.~J.~O.}\ \bibnamefont
  {Reichhardt}},\ }\bibfield  {title} {\enquote {\bibinfo {title} {Reversible
  to irreversible transitions for cyclically driven particles on periodic
  obstacle arrays},}\ }\href {\doibase 10.1063/5.0087916} {\bibfield  {journal}
  {\bibinfo  {journal} {J. Chem. Phys.}\ }\textbf {\bibinfo {volume} {156}},\
  \bibinfo {pages} {124901} (\bibinfo {year} {2022})}\BibitemShut {NoStop}%
\end{thebibliography}%

\end{document}